\documentclass[11pt,twoside]{article}

\def\aap{\mbox{\bf{Astron.\ Astrophys.}}}

\def\apjl{\mbox{\bf{Astrophys.\ J.\ L.}}}

\def\apjs{\mbox{\bf{Astrophys.\ J.\ Suppl.}}}

\newcommand{\Msun}{\mbox{$\mathrm{M}_{\odot}$}}

\usepackage{natbib}
\usepackage{asp2006_selma}
\usepackage{epsf}
\usepackage{psfig}
\usepackage{lscape}
\usepackage{amssymb, amsmath}
\usepackage{graphicx}

\begin{document}
\title{A new evolutionary scenario for the formation of massive
black-hole binaries such as M33 X-7 and IC 10 X-1} \author{S.E. de
Mink$^1$, M. Cantiello$^1$, N. Langer$^{2,1}$, O.R. Pols$^1$ and S.-Ch
Yoon$^{2}$} \affil{$^1$Astronomical Institute Utrecht, Princetonplein
5, 3584 CC Utrecht, The Netherlands, S.E.deMink@uu.nl\\ $^2$Argelander-Institut f\"ur
Astronomie, Auf den H\"ugel 71, 53121 Bonn, Germany}

\begin{abstract}

The formation of close massive black-hole binaries is a challenge for
binary evolutionary models, especially the intriguing system M33 X-7
which harbours one of the most massive stellar-mass black holes (16 \Msun) orbiting a 70 \Msun O-star every 3.5 days. In
standard binary evolution theory an episode of mass transfer or common
envelope is inevitable in a binary with such a small orbital period,
which complicates the formation of a black hole with such a high mass.

To explain this system, we discuss a new binary evolution channel \citep{DeMink+09}, in which rotational mixing plays an important role. In very massive close
binaries,  tides force the rotation rate of the stars to be so high
that rotationally induced mixing becomes very efficient. Helium
produced in the center is mixed throughout the envelope.  Instead of
expanding during their main-sequence evolution (with the inevitable
consequence of mass transfer), these stars stay compact, and avoid filling their Roche lobe.  They gradually  evolve into massive helium stars. This
scenario naturally leads to the formation of very massive black holes
in a very close orbit with a less evolved massive companion such as
M33 X-7.
\end{abstract}

\section{Introduction}

The black holes in the X-ray binaries M33~X-7 and IC10~X-1 are two of
the most massive stellar-mass black holes: $15.65 \pm 1.45\Msun$
\citep{Orosz+07} and 23-34 \Msun
\citep{Prestwich+07,Silverman+Filippenko08} respectively. Such high
masses require that the progenitor star was very massive and
experienced only a moderate mass-loss rate
\citep[e.g.][]{Belczynski+09}. 
However, both black holes orbit a
massive companion star in an close orbit: 3.45 days in the case of M33
X-7 and 1.43 days in the case of IC10~X-1. These orbits are so tight
that radius of the progenitor star must have been larger than the
current separation between the stars.  This implies that the
progenitor experienced severe mass loss via Roche-lobe overflow.  This
is in contradiction with the very moderate mass loss rate required to
achieve such a high mass for the black hole.
Explaining both the high mass of the black hole and the tight orbit simultaneously is a major challenge for binary evolution models. Here, we discuss an alternative evolutionary scenario for very close massive binaries in which mass loss by Roche-lobe overflow is avoided.

\section{Rotational mixing in massive binaries}

%The rotation rate is nowadays considered as one of the main initial
%stellar parameters, along with mass and metallicity, which determine
%the fate of single stars. It can deform the stars, interplay with the
%mass loss and trigger instabilities in the interior leading to
%turbulent mixing in otherwise stable layers \citep[e.g.][]{Maeder+97}.

In models of rapidly rotating, massive stars, rotational mixing can
efficiently transport centrally produced helium throughout the stellar
envelope. Instead of expanding during core H-burning as non-rotating
models do, they stay compact, become more luminous and move blue-wards
in the Hertzsprung-Russell diagram \citep{Maeder87}. This type of
evolution is often referred to as (quasi-)chemically homogeneous
evolution and has been proposed for the formation of long gamma-ray
burst progenitors \citep{Yoon+06,Woosley+06}.

High rotation rates can be readily achieved in binary systems due to
mass and angular momentum transfer \citep{Cantiello+07} and also by tidal interaction in
close binaries \citep{Detmers+08}.  In \citet{DeMink+09} we demonstrated that even in detached, tidally-locked
binaries, rotational mixing can lead to chemically homogeneous
evolution.  In these models it is the less massive star, in which the
effects of rotational mixing are less pronounced, that fills its Roche
lobe first, contrary to what classical binary evolution theory
predicts. In single stars this type of evolution only occurs at low
metallicity, because at solar metallicity mass and angular momentum
loss in the form of a stellar wind spins down the stars and prevents
initially rapidly rotating stars from following nearly chemically
homogeneous evolutionary tracks \citep{Yoon+06, Brott+09}. In a close
binary tides can replenish the angular momentum, opening the
possibility for chemically homogeneous evolution in the solar
neighbourhood.

\begin{figure}[]
% \vspace*{-2.0 cm}
\begin{center}
 \includegraphics[width=\textwidth]{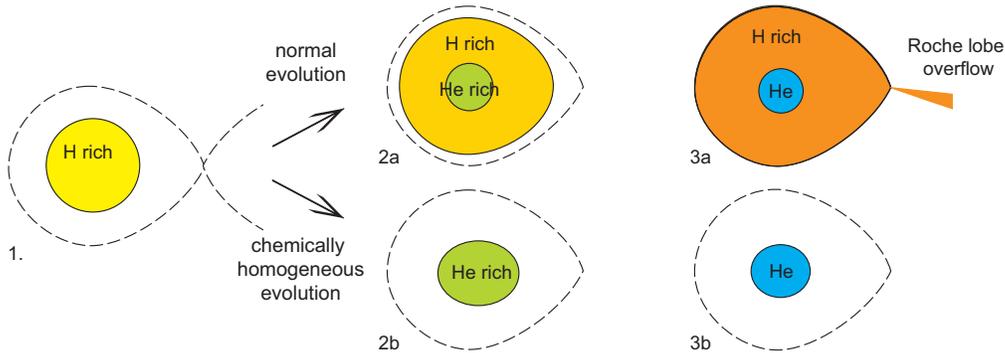} 
% \vspace*{-1.0 cm}
 \caption{Cartoon representation of normal and  chemically homogeneous evolution in a binary system, adapted from \citet{DeMink+08}}
   \label{fig1}
\end{center}
\end{figure}

\section{The formation of short-period black-hole binaries}

The binary models presented by \citet{DeMink+09} all evolve into
contact, due to expansion of the secondary star. However, Roche-lobe
overflow may be avoided altogether in systems in which the secondary
stays compact, either because it also evolves chemically
homogeneously, which may occur if $M_1 \approx M_2$, or because it
evolves on a much longer timescale than the primary, when $M_2\ll
M_1$. Whereas standard binary evolution theory predicts that the
smaller the orbital period, the earlier mass transfer sets in, they
find that binaries with the smallest orbital periods may avoid the
onset of mass transfer altogether, see Fig.~1.  This evolutionary scenario does not
fit in the traditional classification of interacting binaries 
(Case~{\it A}, {\it B} and {\it C}), which is based on the evolutionary stage of
the primary component at the onset of mass transfer
\citep{Kippenhahn+Weigert67,Lauterborn70}.  In the remainder of this
paper we will refer to this new case of binary evolution, in which
mass transfer is delayed or avoided altogether as a result of very
efficient internal mixing, as Case~{\it M}.

The massive and tight systems in which Case~{\it M} can occur are
rare \citep{DeMink+08}. Additional mixing processes induced by the presence of the
companion star, which may be important in such systems, will widen the
parameter space in which Case~{\it M} can occur: it would lower the minimum
mass for the primary star and increase the orbital period below which
this type of evolution occurs.  The massive LMC binary
[L72]~LH~54-425, with an orbital period of 2.25~d
\citep{Williams+08} may be a candidate
for this type of evolution. Another interesting case is the galactic
binary WR20a, which consists of two core hydrogen burning stars of
$82.7\pm5.5$\Msun and $81.9\pm5.5\Msun$ in an orbit of 3.69~d. Both stars
are so compact that they are detached. The surface abundances show
evidence for rotational mixing: a nitrogen abundance of six times
solar, while carbon is depleted \citep{Bonanos+04, Rauw+05}.

%\section{Short-period Wolf-Rayet and black-hole binaries}

If Roche-lobe overflow is avoided throughout the core hydrogen-burning
phase of the primary star, both stars will stay compact while the
primary gradually becomes a helium star and can  be observed as a
Wolf-Rayet star.  Initially the Wolf-Rayet star will be more massive
than its main sequence companion, but mass loss due to the strong
stellar wind may reverse the mass ratio, especially in systems which
started with nearly equal masses. Examples of observed short-period
Wolf-Rayet binaries with a main-sequence companion are
CQ~Cep%
%\footnote{CQ~Cep: ${\rm M_{WR}}= 24\Msun,\,{\rm M_O}=30\Msun,\,P_{\rm orb} = 1.6$~d}
, CX~Cep%
%\footnote{CX~Cep: ${\rm M_{WR}}=20\Msun,\,{\rm M_O}=28\Msun,\, P_{\rm orb} = 2.1$~d}
, HD~193576
%\footnote{HD~193576: ${\rm M_{WR}}=9\Msun,\,{\rm M_O}=29\Msun,\, P_{\rm orb} = 4.2$~d} 
and the very massive system HD~311884
%\footnote{HD~311884: ${\rm M_{WR}}=51\Msun,\,{\rm M_O}=60\Msun,\, P_{\rm orb} = 6.2$~d} 
\citep{vanderHucht01}. Such
systems are thought to be the result of very non-conservative mass
transfer or a common envelope phase
\citep[e.g.][]{Petrovic+05_WR}. Case~M constitutes an alternative
formation scenario which does not involve mass transfer.

Case~{\it M} is particularly interesting for the formation of massive
black-hole binaries, such as M33~X-7 and  IC~10~X-1.
%\footnote{Other examples are
%Cyg~X-1
%%\footnote{Cyg~X-1: ${\rm M_{bh}}\approx 10\Msun,\,{\rm M_O}\approx18\Msun,\,P_{\rm orb}=5.6$~d}
%\citep{Herrero+95}, LMC~X-1
%%\footnote{LMC~X-1: ${\rm M_{bh}} \approx 10\Msun,\,{\rm M_O}\approx30\Msun,\, P_{\rm orb}=3.9$~d}
%\citep{Orosz+08} and LMC~X-3
%%\footnote{LMC~X-3: ${\rm M_{bh}}\approx4$--$10\Msun,\,{\rm M_O}\approx40\Msun,\,P_{\rm orb}=4.2$~ d} 
%\citep[and references therein]{Yao+05}.}. 
The explanation for the
formation of these systems with standard binary evolutionary models or synthetic models  \citep[e.g.][]{Abubekerov+09} involve a common-envelope phase that sets in after the end of core
helium burning, as the progenitor of the black hole
must have had a radius much larger than the current orbital
separation.  This scenario is problematic as it requires the
black-hole progenitor to lose roughly ten times less mass before the
onset of Roche-lobe overflow than what is currently predicted by
stellar evolution models \citep{Orosz+07}.  An additional problem is
that the most likely outcome of the common envelope phase would be a
merger, as the envelopes of massive stars are tightly bound
\citep{Podsiadlowski+03}.  In the Case~{\it M} scenario the black hole
progenitor stays compact and avoids Roche-lobe overflow, at least
until the end of core helium burning.

\section {Conclusion}
We propose an alternative formation scenario for close massive black
hole binaries, such as M33~X-7 and IC10~X-1.  In this scenario the
system starts initially as a very close binary in which tides force
the stars to rotate rapidly. This induces mixing processes in the
progenitor of the black hole, which as a result stays compact within
its Roche lobe. Whereas the short orbital period is a major challenge for classical
binary evolution scenarios, in this scenario it constitutes an
essential ingredient: it results in tidal-locking of the stellar
rotation to the fast revolution of the orbit. The high mass of the
black hole is naturally explained by this scenario: efficient mixing
leads to the formation of very massive helium stars and consequently
massive black holes.

Opportunities to test the validity of this scenario for M33~X-7 may 
come from the properties of the companion star \citep{Valsecchi+09}
and the high Kerr parameter of the black-hole \citep{Liu+08}, which
according to \citet{Mendez09} can only be explained with hypercritical
accretion onto the black-hole.  Further modelling is needed to validate
this statement in the light of this new evolutionary scenario.

%Other examples of stellar mass black-hole binaries with short periods
%and a massive companions, in which nearly chemically homogeneous
%evolution may be important are IC~10~X-1
%\citep{Silverman+Filippenko08}, Cyg~X-1
%%\footnote{Cyg~X-1: ${\rm M_{bh}}\approx 10\Msun,\,{\rm M_O}\approx18\Msun,\,P_{\rm orb}=5.6$~d}
%\citep{Herrero+95}, LMC~X-1
%%\footnote{LMC~X-1: ${\rm M_{bh}} \approx 10\Msun,\,{\rm M_O}\approx30\Msun,\, P_{\rm orb}=3.9$~d}
%\citep{Orosz+08} and LMC~X-3
%%\footnote{LMC~X-3: ${\rm M_{bh}}\approx4$--$10\Msun,\,{\rm M_O}\approx40\Msun,\,P_{\rm orb}=4.2$~ d} 
%\citep[and references therein]{Yao+05}.

%\acknowledgements  
%SdM acknowledges the organizers of the conference and LKBF
%for financial support.

%%% THE BIBLIOGRAPHY
%%%
%%% CONSULT SECTION 3 OF "INSTRUCTIONS FOR AUTHORS" FOR HOW TO USE NATBIB.
%%% AUTHORS ARE ENCOURAGED TO USE EITHER THE "THEBIBLIOGRAPY" ENVIRONMENT
%%% BY UNCOMMENTING (DELETING THE "%" SYMBOL) THE COMMANDS BELOW, OR BY
%%% USING THE BIBTEX ENVIRONMENT. TO FIND OUT WHICH IS APPLICABLE TO YOUR
%%% CONTRIBUTION, CONSULT THE VOLUME EDITORS FOR YOUR PROCEEDINGS.
%%%

%\begin{thebibliography}{}
\bibliographystyle{rjr-asp-bib}
\bibliography{references}

\begin{thebibliography}{}

\bibitem[\protect\astroncite{{Abubekerov} et~al.}{2009}]{Abubekerov+09}
{Abubekerov} M.~K., {Antokhina} E.~A., {Bogomazov} A.~I., {Cherepashchuk}
  A.~M., 2009,
  Astronomy Reports\   53, 232

\bibitem[\protect\astroncite{{Belczynski} et~al.}{2009}]{Belczynski+09}
{Belczynski} K., {Bulik} T., {Fryer} C.~L., {Ruiter} A., {Vink} J.~S., {Hurley}
  J.~R., 2009,
  ArXiv/0904.2784\

\bibitem[\protect\astroncite{{Bonanos} et~al.}{2004}]{Bonanos+04}
{Bonanos} A.~Z., {Stanek} K.~Z., {Udalski} A., {Wyrzykowski} L.,
  {{\.Z}ebru{\'n}} K., {Kubiak} M., {Szyma{\'n}ski} M.~K., {Szewczyk} O.,
  {Pietrzy{\'n}ski} G., {Soszy{\'n}ski} I., 2004,
  \apjl\   611, L33

\bibitem[\protect\astroncite{{Brott} \& {et al.}}{2009}]{Brott+09}
{Brott} I., {et al.}, 2009,
  in prep.\

\bibitem[\protect\astroncite{{Cantiello} et~al.}{2007}]{Cantiello+07}
{Cantiello} M., {Yoon} S.-C., {Langer} N., {Livio} M., 2007,
  \aap\   465, L29

\bibitem[\protect\astroncite{{De Mink} et~al.}{2009}]{DeMink+09}
{De Mink} S.~E., {Cantiello} M., {Langer} N., {Pols} O.~R., {Brott} I., {Yoon}
  S.-C., 2009,
  \aap\   497, 243

\bibitem[\protect\astroncite{{De Mink} et~al.}{2008}]{DeMink+08}
{De Mink} S.~E., {Cantiello} M., {Langer} N., {Yoon} S.-C., {Brott} I.,
  {Glebbeek} E., {Verkoulen} M., {Pols} O.~R., 2008,
\newblock in L. {Deng}, K.~L. {Chan} (eds.), IAU Symposium,
  Vol. 252 of {\em IAU Symposium\/},  365

\bibitem[\protect\astroncite{{Detmers} et~al.}{2008}]{Detmers+08}
{Detmers} R.~G., {Langer} N., {Podsiadlowski} P., {Izzard} R.~G., 2008,
  \aap\   484, 831

\bibitem[\protect\astroncite{{Kippenhahn} \&
  {Weigert}}{1967}]{Kippenhahn+Weigert67}
{Kippenhahn} R., {Weigert} A., 1967,
  Zeitschrift fur Astrophysik\   65, 251

\bibitem[\protect\astroncite{{Lauterborn}}{1970}]{Lauterborn70}
{Lauterborn} D., 1970,
  \aap\   7, 150

\bibitem[\protect\astroncite{{Liu} et~al.}{2008}]{Liu+08}
{Liu} J., {McClintock} J.~E., {Narayan} R., {Davis} S.~W., {Orosz} J.~A., 2008,
  \apjl\   679, L37

\bibitem[\protect\astroncite{{Maeder}}{1987}]{Maeder87}
{Maeder} A., 1987,
  \aap\   178, 159

\bibitem[\protect\astroncite{{M{\'e}ndez}}{2009}]{Mendez09}
{M{\'e}ndez} E.~M., 2009,
\newblock in C. {Meegan}, C. {Kouveliotou}, N. {Gehrels} (eds.), American
  Institute of Physics Conference Series,
  Vol. 1133 of {\em American Institute of Physics Conference Series\/},  109

\bibitem[\protect\astroncite{{Orosz} et~al.}{2007}]{Orosz+07}
{Orosz} J.~A., {McClintock} J.~E., {Narayan} R., {Bailyn} C.~D., {Hartman}
  J.~D., {Macri} L., {Liu} J., {Pietsch} W., {Remillard} R.~A., {Shporer} A.,
  {Mazeh} T., 2007,
  \nat\   449, 872

\bibitem[\protect\astroncite{{Petrovic} et~al.}{2005}]{Petrovic+05_WR}
{Petrovic} J., {Langer} N., {van der Hucht} K.~A., 2005,
  \aap\   435, 1013

\bibitem[\protect\astroncite{{Podsiadlowski} et~al.}{2003}]{Podsiadlowski+03}
{Podsiadlowski} P., {Rappaport} S., {Han} Z., 2003,
  \mnras\   341, 385

\bibitem[\protect\astroncite{{Prestwich} et~al.}{2007}]{Prestwich+07}
{Prestwich} A.~H., {Kilgard} R., {Crowther} P.~A., {Carpano} S., {Pollock}
  A.~M.~T., {Zezas} A., {Saar} S.~H., {Roberts} T.~P., {Ward} M.~J., 2007,
  \apjl\   669, L21

\bibitem[\protect\astroncite{{Rauw} et~al.}{2005}]{Rauw+05}
{Rauw} G., {Crowther} P.~A., {De Becker} M., {Gosset} E., {Naz{\'e}} Y., {Sana}
  H., {van der Hucht} K.~A., {Vreux} J.-M., {Williams} P.~M., 2005,
  \aap\   432, 985

\bibitem[\protect\astroncite{{Silverman} \&
  {Filippenko}}{2008}]{Silverman+Filippenko08}
{Silverman} J.~M., {Filippenko} A.~V., 2008,
  \apjl\   678, L17

\bibitem[\protect\astroncite{{Valsecchi} et~al.}{2009}]{Valsecchi+09}
{Valsecchi} F., {Willems} B., {Fragos} T., {Kalogera} V., 2009,
  ArXiv/0902.3700\

\bibitem[\protect\astroncite{{van der Hucht}}{2001}]{vanderHucht01}
{van der Hucht} K.~A., 2001,
  New Astronomy Review\   45, 135

\bibitem[\protect\astroncite{{Williams} et~al.}{2008}]{Williams+08}
{Williams} S.~J., {Gies} D.~R., {Henry} T.~J., et al., 2008,
  \apj\   682, 492

\bibitem[\protect\astroncite{{Woosley} \& {Heger}}{2006}]{Woosley+06}
{Woosley} S.~E., {Heger} A., 2006,
  \apj\   637, 914

\bibitem[\protect\astroncite{{Yoon} et~al.}{2006}]{Yoon+06}
{Yoon} S.-C., {Langer} N., {Norman} C., 2006,
  \aap\   460, 199

\end{thebibliography}

%\bibitem[]{}
%\end{thebibliography}

\end{document}